\documentclass[twocolumn,prb,superscriptaddress]{revtex4}
\usepackage{graphicx}
\usepackage{dcolumn}
\usepackage{bm}
\usepackage{amsmath}
\usepackage{times}
\usepackage{epstopdf}
\usepackage{color}
\usepackage[breaklinks=true,colorlinks,citecolor=blue,linkcolor=blue,urlcolor=blue]{hyperref}

\makeatletter

\newcommand{\Rmnum}[1]{\expandafter\@slowromancap\romannumeral #1@}
\makeatother
\graphicspath{{images/}}

\begin{document}

\author{Mahammad Tahir}
\affiliation{Department of Physics, Indian Institute of Technology Kanpur, Kanpur- 208016, India}
\author{Dhananjay Tiwari}
\affiliation{Advanced Safety and User Experience, Aptiv Services Poland, Krakow 30-707, Poland}
\author{Abhishek Juyal}
\affiliation{Georgia Tech Lorraine IRL 2958 - CNRS 57070 Metz, France}
\affiliation{School of Physics, Georgia Institute of Technology, Atlanta, Georgia 30332, USA}
\author{Rohit Medwal}
\affiliation{Department of Physics, Indian Institute of Technology Kanpur, Kanpur- 208016, India}
\author{Soumik Mukhopadhyay}
\affiliation{Department of Physics, Indian Institute of Technology Kanpur, Kanpur- 208016, India}
\title{Interfacial magnetic anisotropy controlled spin pumping in {Co$_{60}$Fe$_{20}$B$_{20}$}/{Pt} stack} 

\begin{abstract}
Controlled spin transport in magnetic stacks is required to realize pure spin current-driven logic and memory devices. The control over the generation and detection of the pure spin current is achieved by tuning the spin to charge conversion efficiency of the heavy metal interfacing with ferromagnets. Here, we demonstrate the direct tunability of spin angular momentum transfer and thereby spin pumping, in CoFeB/Pt stack, with interfacial magnetic anisotropy. The ultra-low thickness of CoFeB thin film tilts the magnetic easy axis from in-plane to out-of-plane due to surface anisotropy. The Ferromagnetic resonance measurements are performed to investigate the magnetic anisotropy and spin pumping in CoFeB/Pt stacks. We clearly observe tunable spin pumping effect in the CoFeB/Pt stacks with varying CoFeB thicknesses. The spin current density, with varying ferromagnetic layer thickness, is found to increase from 0.11 to 0.24 MA/m$^2$, with increasing in-plane anisotropy field. Such interfacial anisotropy-controlled generation of pure spin current can potentially lead to next-generation anisotropic spin current-controlled spintronic devices.

\end{abstract}

\maketitle

\section{Introduction}

The direct control of the generation and detection of pure spin current is essential for next-generation nanoscale spin devices such as spin wave logic, spin orbit torque induced magnetization switching, and spin Hall nano oscillators~\cite{Dieny,Brataas,Jungwirth}. Spintronic devices which utilize spin transport across interfaces composed of a ferromagnetic (FM) and a nonmagnetic (NM) layer with large spin-orbit interaction (SOI) are promising devices for the spin-to-charge conversion for future applications~\cite{Bader, Tser, Sanchez, Tao, Uchida, Avci}. Pure spin current has numerous advantages over conventional charge current owing to its low power consumption, negligible Joule heating, and absence of stray fields~\cite{Chappert, Puebla}. At ferromagnetic resonance (FMR) condition, the spin pumping effect allows injection of a pure spin current from the FM into the NM layer~\cite{Tser2,Tser3}. Spin pumping can be observed by detecting enhanced damping generated by a spin angular momentum outflow from the source FM to NM layer~\cite{Urban,Tser2}. The enhancement of Gilbert damping is more prominent in NM layers with high spin orbit coupling (SOC) because of stronger interaction between electron spin and lattice. The generated pure spin current is converted into a charge current via inverse spin Hall effect (ISHE) in heavy metal (HM) utilizing the SOC~\cite{Saitoha,Mosendz}.

\begin{figure}
\includegraphics[width=0.99\linewidth]{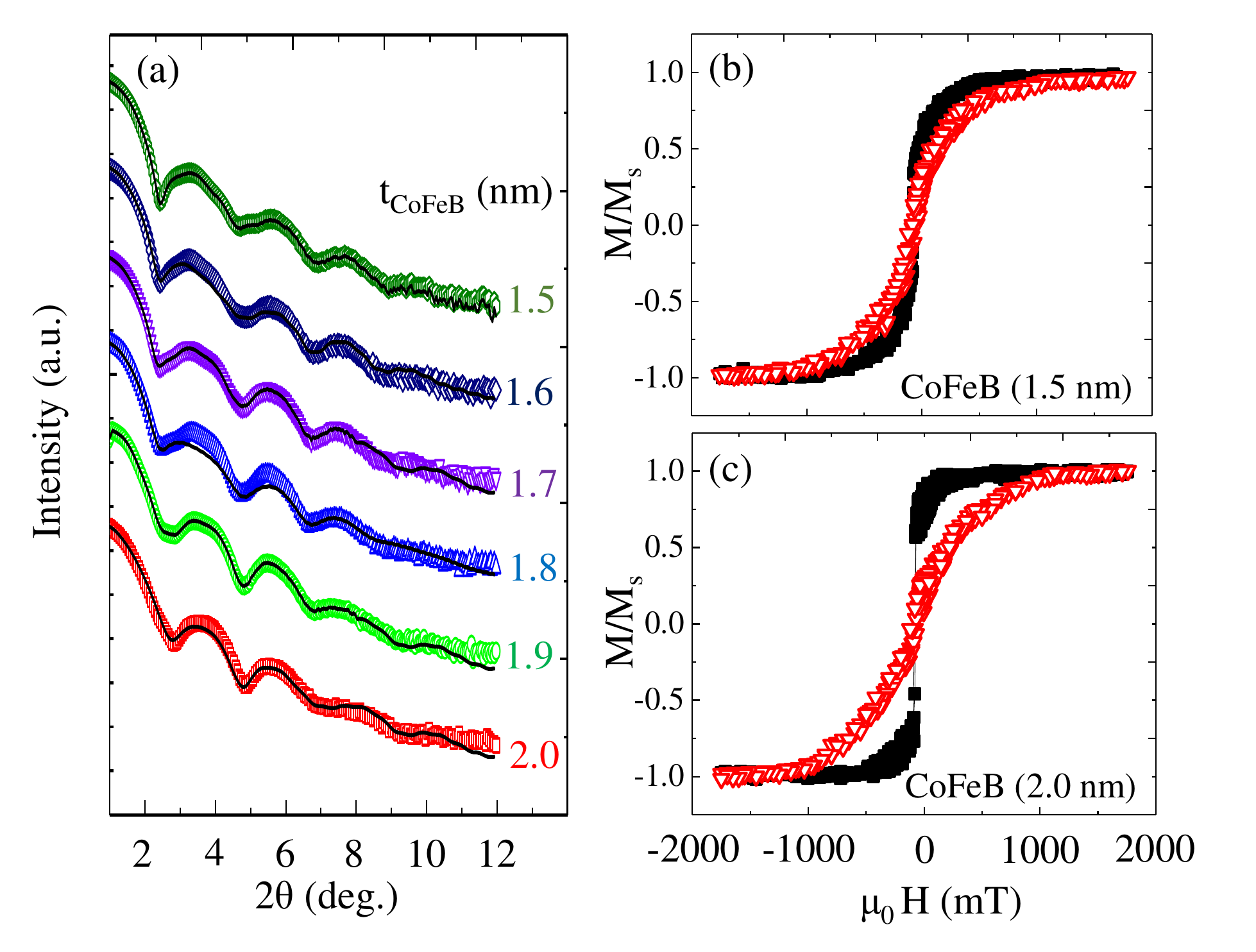}
\caption{(a) XRR spectra of Pt ($4$ nm)/CoFeB ($1.5$–$2.0$ nm)/Al ($4$ nm) stacks. Open symbols are experimental data points while solid lines represent simulations. (b), (c) In-plane (solid black squares) and out-of-plane (open red triangles) magnetization hysteresis loops for the stacks involving CoFeB(1.5 nm) and CoFeB(2.0 nm), respectively.}
\label{fig1}
\end{figure}

\begin{figure}
\includegraphics[width=0.99\linewidth]{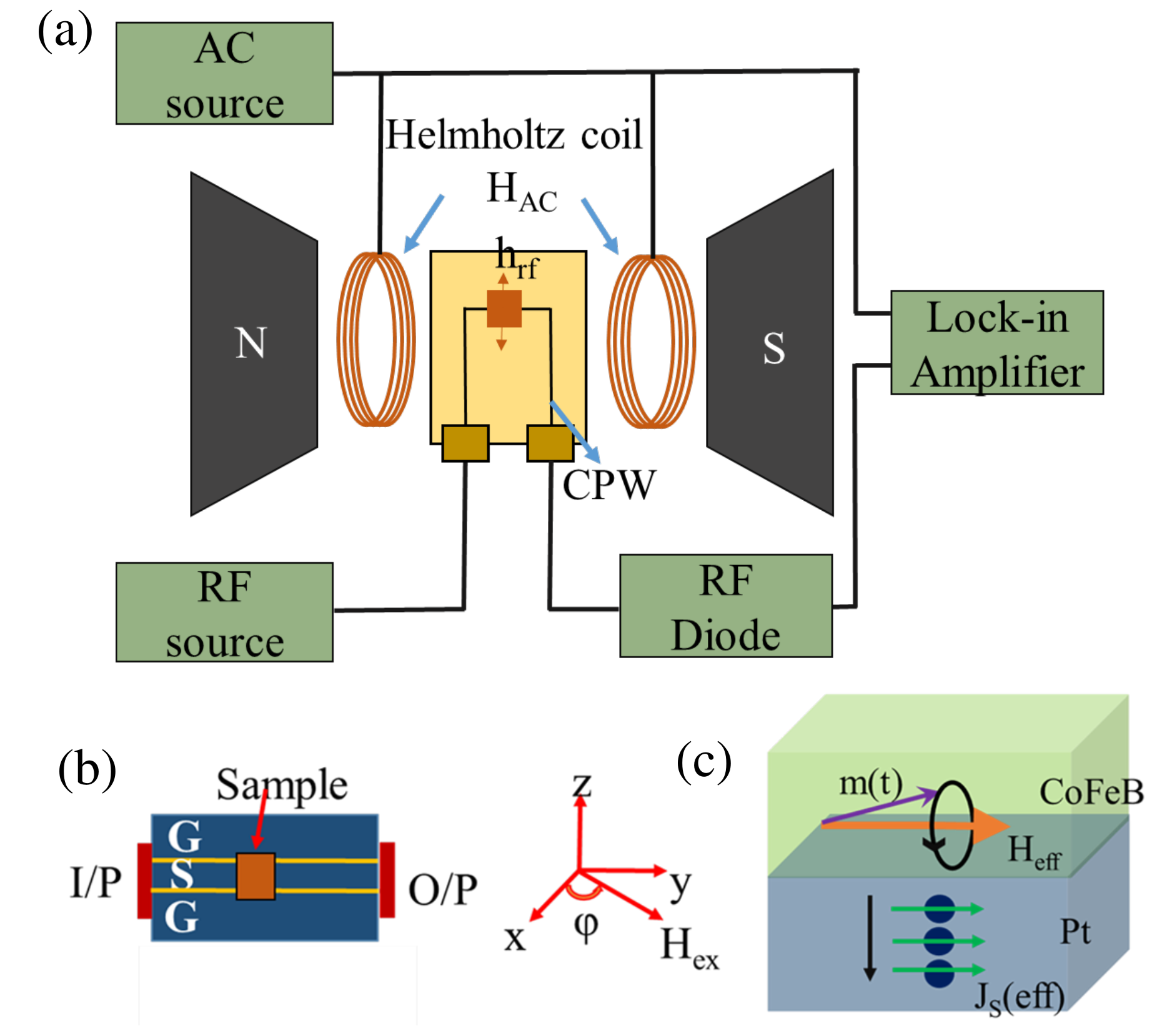}
\caption{(a) The FMR measurement configuration is shown schematically. (b) Schematic of CPW showing the thin film stack in contact with the main signal transmission line (S), which is isolated from the adjacent ground line (G). The input and output signal ports of the CPW are represented by I/P and O/P. (c) A schematic of the CoFeB/Pt stack is shown to demonstrate how spin pumping induces the generation and flow of spin current $\mathrm{J}_{\mathrm{s}}$ across the CoFeB/Pt interface.}
\label{fig2}
\end{figure}
\begin{figure*}
\includegraphics[width=0.9\linewidth]{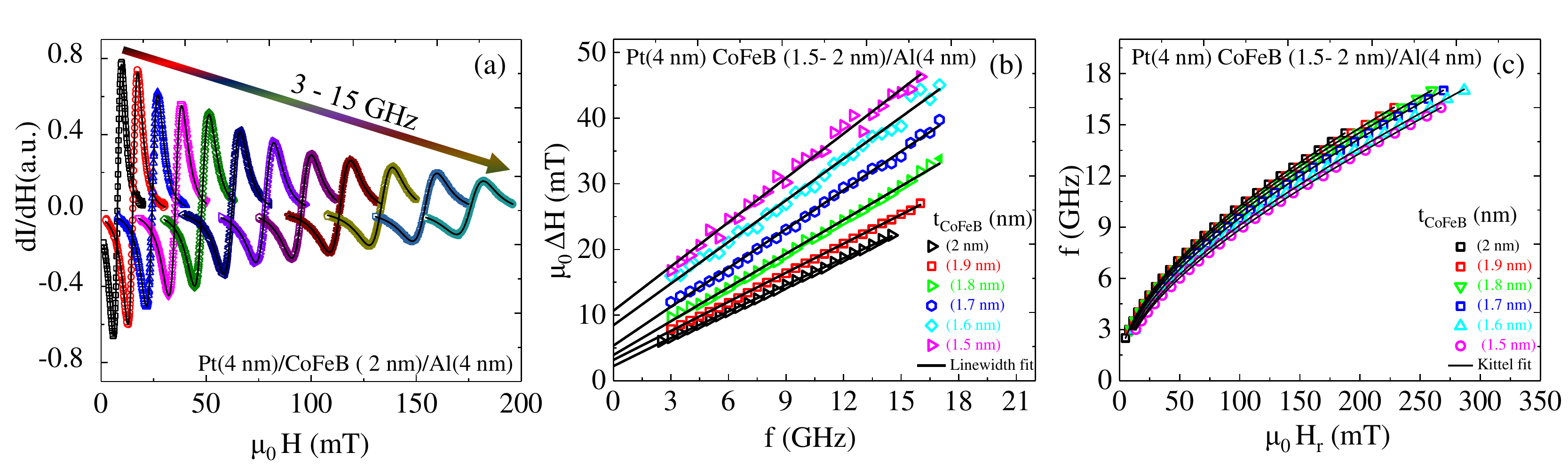}
\caption{(a) FMR spectra recorded on a Pt(4 nm)/CoFeB(2 nm)/Al(4 nm) stack. Open symbols are experimental data points and black solid lines show the fitting using Eq.~\ref{eq1}. (b) Linewidth ($\mu_{0}\Delta \mathrm{H}$) versus frequency ($\mathrm{f}$) for all the samples. Black continuous lines are the linear fits as described by Eq.~\ref{eq3}. (c) Resonance field $\mu_{0}\mathrm{H}_{\mathrm{r}}$ vs $\mathrm{f}$ for all the samples. Black lines are the fits using Eq.~\ref{eq2}.}
\label{fig3}
\end{figure*}
The control over the generation and detection of the pure spin current is achieved by material engineering of HMs with enhanced SOC interfacing with FM ~\cite{Wang}. The most studied HMs with large SOC explored via spin pumping are the 4d and 5d metals, such as Pt~\cite{A,Z}, Pd~\cite{Dong,Kumar}, Mo~\cite{Husain}, $\beta$-W~\cite{Panda}, $\beta$-Ta~\cite{Akansel}. The direction-sensitive flow of pure spin current can possibly be realized by controlling (i) the magnetic anisotropy of the spin sink layer and relative alignment of the two magnetic layers in the FM/spacer/FM structure~\cite{Baker} and (ii) the crystalline symmetry of the ferromagnet where anisotropic Gilbert damping can be tuned by magnetization orientation~\cite{Li, Rohit}. This enables the anisotropic generation and relaxation of pure spin current suitable for direction-specific propagation of spin current~\cite{Tser}.

Similar control over the pure spin current can be achieved by the shape and size effect of FM and HM. The size effect provides an additional interfacial anisotropy to influence the spin dynamics and thus, spin pumping. The interface anisotropy can largely affect two processes (i) spin current transmission at the interface defined by the spin mixing conductance ($\mathrm{g}^{\uparrow\downarrow}_{\mathrm{eff}}$) and (ii) spin relaxation at the interface. Spin mixing conductance ($\mathrm{g}^{\uparrow\downarrow}_{\mathrm{eff}}$) depends on the quality of the interface between FM and HM, the thickness of the FM and the type of HMs used as spin sink~\cite{Tser2,Tser3,Nakayama}. As CoFeB possesses remarkable magnetic properties viz. low Gilbert damping factor, low uniaxial anisotropy due to its amorphous nature and spin polarization, it is particularly suitable for MRAM application~\cite{Huang}. It is, therefore, extremely important to study the spin dynamics of the CoFeB/Pt interface. 

In the present work, we demonstrate interfacial anisotropy dependent generation and relaxation of spin current. The CoFeB thin films of varying thickness are fabricated to control the magnetic anisotropy and probe the corresponding influence on the spin pumping effect. The ultra low thickness of the films tilts the magnetic easy axis of the ferromagnetic layer from in-plane to out-of-plane. We investigate the effect of the thickness of the ferromagnetic layer and thereby the interfacial anisotropy fields on spin pumping using broadband FMR spectroscopy.

\section{Experimental Methods}

In order to investigate the interfacial anisotropic spin pumping, Si/SiO$_{2}$/Pt(4 nm)/Co$_{60}$Fe$_{20}$B$_{20}$(t$_{\mathrm{CoFeB}}$)/Al(4 nm) stacks are fabricated using DC magnetron sputtering with CoFeB ferromagnetic layer’s thickness varying from $\mathrm{t}_{\mathrm{CoFeB}} = $ 1.5 nm to $\mathrm{t}_{\mathrm{CoFeB}} = $ 2.0 nm with step size of 0.1 nm. The target is pre-sputtered for two minutes to avoid any contamination during the deposition. All thin films are deposited at a constant deposition rate of 0.8 \AA/sec. The control samples, without 4.0 nm Pt layer, are also deposited under the same condition. To prevent surface oxidation of the thin CoFeB layer, 4 nm Al layer is deposited as a capping layer. To check the quality of samples, the deposited stacks are subjected to X-ray reflectivity (XRR) measurements for accurate estimation of the thickness and interface roughness using Smartlab Rigaku X-ray diffractometer. In order to investigate the magnetization dynamics and magnetic anisotropy of the stacks, lock-in based broadband ferromagnetic resonance (FMR) technique (NanOsc) is used. The 4 × 2 mm$^2$ sample is placed on a 200 - $\mu$m wide-coplaner waveguide (CPW) in flip-chip manner. FMR measurements are carried out in the broad frequency range (3–15 GHz) as a function of in-plane external DC magnetic field at room temperature. 

\section{Results and Discussion}

The XRR spectra are recorded on the stacks to determine their exact thickness, density, and interface roughness. The observable Kiessig fringes throughout the whole measurement range confirms the excellent quality of the films and their interfaces (Fig.~\ref{fig1}a). The simulation of the recorded spectra is done using X-ray reflectivity software (segmented V.1.2) considering a stack of Si/SiO$_{2}$/Pt/Co$_{60}$Fe$_{20}$B$_{20}$/Al as shown in Fig.~\ref{fig1}a. The low interface roughness of the films and the fact that each layer's thickness approximately matches the value obtained from deposition rate calibration indicates that there is no intermixing of CoFeB and Pt in the fabrication of our films. Spin transport channels are affected by a rough FM/NM interface since it is related to the spin impedance matching as well as SOC of the interface ~\cite{Jin}. The estimated values for each layer's thickness, density, and interface roughness are shown in Table~\ref{table1}.

\begin{table*}
        \caption{The simulated values of density, thickness and roughness of each layer of Si/SiO$_{2}$/Pt(4 nm)/Co$_{60}$Fe$_{20}$B$_{20}$(t$_{\mathrm{CoFeB}}$)/Al(4 nm) stacks are shown}\label{table1}
        \scalebox{0.75}{
    \begin{tabular}{@{\extracolsep{8pt}}lccccccccc@{}} \hline
        \multicolumn{1}{l}{} &
        \multicolumn{3}{c}{Density ( gm/cm$^3$)} &
        \multicolumn{3}{c}{Thickness (nm)} &
        \multicolumn{3}{c}{Interface Roughnesses (nm)} \\\\
        \cline{2-4}
        \cline{5-7}
        \cline{8-10}

Sample              & CoFeB       & Pt   & Al       & CoFeB          & Pt    & Al       & Substrate/Pt      & Pt/CoFeB  & CoFeB/Al \\ 

\hline
            &               &               &                   &           &           &           &       &   &\\

CoFeB(2 nm)   &      7.95$\pm$0.84   &      22.45$\pm$0.94   &       2.45 $\pm$0.62   &       2.01$\pm$0.04   &      3.99$\pm$0.05  &       3.33$\pm$0.52   &      0.33$\pm$0.02   &        0.34$\pm$0.04   &      0.25$\pm$0.06   \\
   \\        
CoFeB(1.9 nm)        &       7.98$\pm$0.74  &      21.95$\pm$0.76   &      2.94$\pm$0.42   &       1.92$\pm$0.28&       4.02$\pm$0.02&       3.25$\pm$0.65&      0.28$\pm$0.03 &        0.38$\pm$0.06   &      0.23$\pm$0.07  \\
 \\         
CoFeB(1.8 nm)        &        8.15 $\pm$0.94   &       22.11 $\pm$0.68   &        2.82$\pm$0.52   &        1.78 $\pm$0.06   &      3.96$\pm$0.06 &       3.25$\pm$0.65   &       0.38 $\pm$0.04  &      0.32$\pm$0.06   &        0.26$\pm$0.04   \\
 \\
CoFeB(1.7 nm)    &        8.68$\pm$0.68     &      21.50$\pm$0.56  &      2.12 $\pm$0.64  &       1.65$\pm$0.03  &      4.01$\pm$0.04   &      3.45$\pm$0.53   &       0.30 $\pm$0.04   &       0.34$\pm$0.05    &      0.46$\pm$0.04    \\
   \\

CoFeB(1.6 nm)    &        7.98$\pm$0.48    &      22.24$\pm$0.78  &      2.83$\pm$0.33 &       1.53$\pm$0.08  &      3.95$\pm$0.06   &      3.60$\pm$0.48   &       0.35$\pm$0.02    &       0.32$\pm$0.05   &      0.50$\pm$0.06   \\
       \\

CoFeB(1.5 nm)      &        7.85 $\pm$0.74    &      22.85 $\pm$0.88  &      2.95$\pm$0.56 &       1.46$\pm$0.04  &      3.90$\pm$0.04    &      3.51$\pm$0.59   &       0.38$\pm$0.06   &       0.33$\pm$0.08   &      0.22$\pm$0.06   \\
\\
            
            \\ \\ \hline 
    \end{tabular}}
\end{table*}

\begin{figure}
\includegraphics[width=0.8\linewidth]{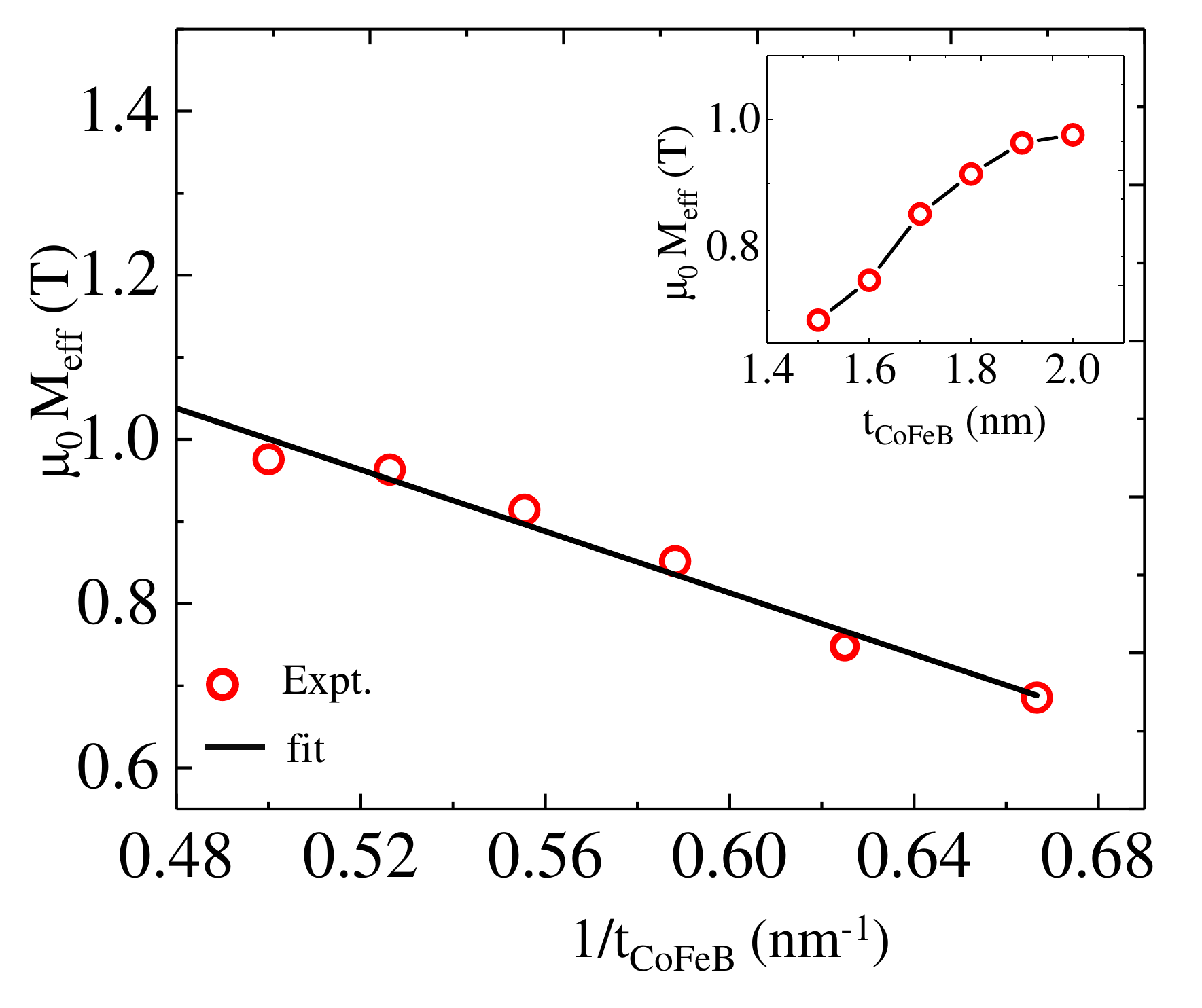}
\caption{$\mu_{0}\mathrm{M}_{\mathrm{eff}}$ versus inverse of thickness (1/$\mathrm{t}_{\mathrm{CoFeB}}$) of the CoFeB layer. Inset shows the variation of $\mu_{0}\mathrm{M}_{\mathrm{eff}}$ versus thickness of the CoFeB ($\mathrm{t}_{\mathrm{CoFeB}}$) layer}.
\label{fig4}
\end{figure}
\begin{figure*}
\includegraphics[width=0.99\linewidth]{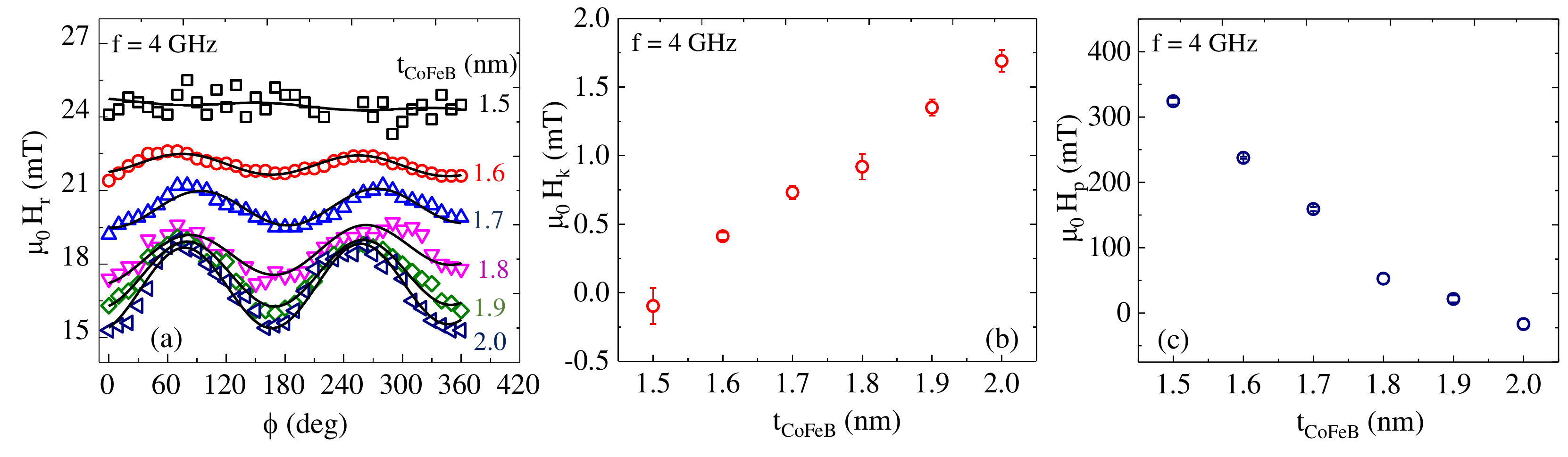}
\caption{(a) Azimuthal angle $\phi$ dependence of $\mu_{0}\mathrm{H}_{\mathrm{r}}$ for f = 4 GHz. Sinusoidal dependence shows the presence of an in-plane anisotropy in addition to the perpendicular magnetic anisotropy. The experimental data are fitted (solid line) with Eq.~\ref{eq5} to obtain $\mu_{0}\mathrm{H}_{\mathrm{k}}$ and $\mu_{0}\mathrm{H}_{\mathrm{p}}$. (b) and (c) shows the variation of in plane magnetic anisotropy $\mu_{0}\mathrm{H}_{\mathrm{k}}$  and out of plane magnetic anisotropy $\mu_{0}\mathrm{H}_{\mathrm{p}}$ with thickness of CoFeB layer ($\mathrm{t}_{\mathrm{CoFeB}}$).}
\label{fig5}
\end{figure*}

The typical in-plane (solid black squares) and out-of-plane (open red triangles) magnetic hysteresis loops for  Pt/Co$_{60}$Fe$_{20}$B$_{20}$/Al stacks are shown in Fig.~\ref{fig1}b and Fig.~\ref{fig1}c. The hysteresis loops for $\mathrm{t}_{\mathrm{CoFeB}}=2$ nm layer suggests existence of in-plane magnetic anisotropy of the CoFeB layer. As the Pt layer thickness is 4 nm, the PMA occurs due to the increased dominance of d–d hybridization effect at the CoFeB/Pt interface when the CoFeB layer thickness is reduced to 1.5 nm. The perpendicular easy axis emerges as a consequence of the increased interfacial magnetic anisotropy and decreased demagnetization energy with decreasing CoFeB thickness. The interfacial PMA stack in our case is composed of a thin ferromagnetic layer (CoFeB) sandwiched between a heavy metal layer (Pt) and a capping layer (Al). The heavy metal layer (Pt) interface with the ferromagnetic layer is responsible for the spin Hall effect, which is potentially useful for SOT and skyrmion based devices~\cite{Guoqiang}.
\begin{figure*}
\includegraphics[width=0.99\linewidth]{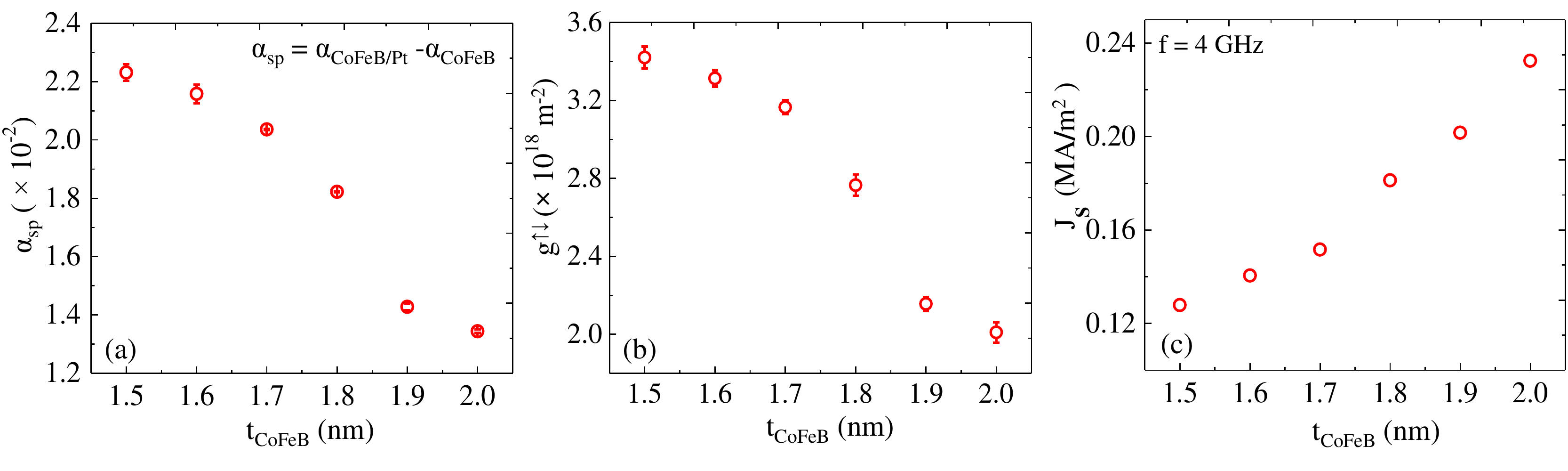}
\caption{(a) Spin pumping induced Gilbert damping factor $\alpha_{\mathrm{sp}}$ with respect to the thickness ($\mathrm{t}_{\mathrm{CoFeB}}$) of CoFeB layer.(b) Effective interfacial spin-mixing conductance $\mathrm{g}^{\uparrow\downarrow}_{\mathrm{eff}}$ vs $\mathrm{t}_{\mathrm{CoFeB}}$ of the CoFeB(1.5 - 2 nm)/Pt stacks evaluated using  Eq.~\ref{eqn6}.(c) Effective spin-current density (injected in Pt) vs $\mathrm{t}_{\mathrm{CoFeB}}$ calculated at $\mathrm{f} = $ 4 GHz using Eq.~\ref{eq7}.}
\label{fig 6}
\end{figure*}

Fig.~\ref{fig2}a shows the measurement configuration of broadband FMR spectrometer. The derivative Lorentzian function (Eq.~\ref{eq1}) having symmetric and asymmetric coefficients is used to fit the recorded FMR spectra of Pt(4 nm)/Co$_{60}$Fe$_{20}$B$_{20}$(t$_{\mathrm{CoFeB}}$)/Al(4 nm) stack (Fig.~\ref{fig3}a) ~\cite{Woltersdorf}.
\begin{widetext}
\begin{equation}
\frac{d \mathrm{I}_{\mathrm{FMR}}}{d \mathrm{H}} = 4\mathrm{A}\frac{\Delta \mathrm{H}( \mathrm{H}- \mathrm{H}_{\mathrm{r}})}{(4( \mathrm{H}- \mathrm{H}_{\mathrm{r}})^2+
(\Delta  \mathrm{H})^2)^2}-\mathrm{S}\frac{(\Delta  \mathrm{H})^2-4(H-H_{\mathrm{r}})^2}{(4( \mathrm{H}- \mathrm{H}_{\mathrm{r}})^2+(\Delta  \mathrm{H})^2)^2}\label{eq1}
\end{equation}
\end{widetext}
where $\mathrm{H}$, $\mu_{0}\Delta \mathrm{H}$ and $\mu_{0} \mathrm{H}_{\mathrm{r}}$ are the in plane applied DC magnetic field, FMR line-width and resonance field of microwave absorption, respectively. The FMR signal's amplitudes $\mathrm{S}$ and $\mathrm{A}$ are symmetric and antisymmetric, respectively~\cite{Woltersdorf}. We evaluate the $\mu_{0}\Delta \mathrm{H}$ and $\mu_{0} \mathrm{H}_{\mathrm{r}}$ from the fittings of FMR spectra. The plots of  $\mu_{0}\Delta \mathrm{H}$ versus microwave absorption frequency $\mathrm{f}$ and $\mu_{0} \mathrm{H}_{\mathrm{r}}$ versus $\mathrm{f}$ are shown in Fig.~\ref{fig3}b and Fig.~\ref{fig3}c, respectively. The effective magnetization $\mu_{0} \mathrm{M}_{\mathrm{eff}}$ is calculated as function of CoFeB thickness using Kittel’s equation (Eq.~\ref{eq2})~\cite{Kittel}. 
\begin{equation}
\mathrm{f} = \frac{\mu_{0}\gamma}{2\pi}\left[\left( \mathrm{H}_{\mathrm{r}}+ \mathrm{H}_{\mathrm{k}}\right) \left( \mathrm{H}_{\mathrm{r}}+ \mathrm{H}_{\mathrm{k}}+  \mathrm{M}_{\mathrm{eff}}\right)\right]^{1/2}\label{eq2}
\end{equation}
We evaluate the gyromagnetic ratio ($\gamma=\frac{ \mathrm{g}\mu_{B}}{\hbar} =  1.94 \times$10$^{2}$ GHz$/$T), where $ \mathrm{g}$ is the spectroscopic Lande g factor, $\hbar$ is the reduced planck constant and $\mu_{0} \mathrm{H}_{\mathrm{k}}$ is the in plane anisotropy field  of the FM layer, respectively. The effective Gilbert damping parameter, ($\alpha_{\mathrm{eff}}$) is evaluated by fitting the data shown in Fig.~\ref{fig3}b using the following damping equation~\cite{Kittel}:
\begin{eqnarray}
\mu_{0}\Delta \mathrm{H}=\frac{4\pi \alpha_{\mathrm{eff}}}{\gamma} \mathrm{f}+\mu_{0}\Delta\mathrm{H}_0 \label{eq3}
\end{eqnarray}
Where $\mu_{0}\Delta\mathrm{H}_0$ is the inhomogeneous line-width broadening, related to the homogeneity of the thin film. The linear behaviour of $\mu_{0}\Delta \mathrm{H}$ versus $\mathrm{f}$ confirms the intrinsic origin of damping parameter observed in our Pt(4 nm)/Co$_{60}$Fe$_{20}$B$_{20}$(t$_{\mathrm{CoFeB}}$)/Al(4 nm) stacks as shown in Fig.~\ref{fig3}b. The value of $\alpha_{\mathrm{eff}}$ can also be enhanced by intrinsic and several other extrinsic contributions: 
\begin{eqnarray*}
\alpha_{\mathrm{eff}} = \alpha_{\mathrm{int}}+\alpha_{\mathrm{imp}}+\alpha_{\mathrm{MPE}}+\alpha_{\mathrm{sp}}
\end{eqnarray*}
where $\alpha _{\mathrm{int}}$ is the intrinsic damping parameter, characteristic of the material under investigation (film growth conditions may, however, influence it strongly) and it is the sum of the losses by magnon-magnon scattering and by energy transfer to the lattice, while $\alpha_{\mathrm{imp}}$, $\alpha_{\mathrm{MPE}}$ and $\alpha _{\mathrm{sp}}$ are the extrinsic contributions due to the impurity effects, magnetic proximity effects such as contribution of dynamic coupling between ordered spins in Pt, and spin pumping ~\cite{Conca}, respectively. The variation of $\mu_{0} \mathrm{M}_{\mathrm{eff}}$ as a function of the inverse of thickness of the CoFeB layer is shown in Fig.~\ref{fig4}. In order to estimate the value of saturation magnetization, Eq.~\ref{eq4} is utilised to fit the CoFeB thickness dependence of the effective magnetization~\cite{Kittel}. The inset of Fig.~\ref{fig4} shows the variation of $\mu_{0} \mathrm{M}_{\mathrm{eff}}$ with the thickness of the CoFeB layer. 
\begin{equation}
\mu_{0} \mathrm{M}_{\mathrm{eff}} = \mu_{0} \mathrm{M}_{\mathrm{s}} + \frac{2\mathrm{K}_{\mathrm{s}}}{\mathrm{M}_{\mathrm{s}}\mathrm{t}_{\mathrm{CoFeB}}}\label{eq4}
\end{equation} 
where $\mathrm{t}_{\mathrm{CoFeB}}$, $\mu_{0}\mathrm{M}_{\mathrm{s}}$, $\mathrm{K}_{\mathrm{s}}$, are the thickness of CoFeB layer, saturation magnetization, perpendicular surface magnetic anisotropy of the FM layer, respectively. (Fig.~\ref{figA2}a in the Appendix describes how the effective Gilbert damping parameter, $\alpha_{\mathrm{eff}}$, is influenced by the FM layer's thickness, with the thickness of NM layer being constant.) 

In order to estimate the in-plane anisotropy field $\mu_{0}\mathrm{H}_{\mathrm{k}}$ (IP) and the out-of-plane anisotropy field  $\mu_{0}\mathrm{H}_{\mathrm{p}}$ (OOP), in-plane FMR spectra for the Pt(4 nm)/Co$_{60}$Fe$_{20}$B$_{20}$ (t$_{\mathrm{CoFeB}}$)/ Al(4 nm) stacks are recorded at different azimuthal angles $\phi$ that range from $0^{\circ}$ to $360^{\circ}$. The resonance field $\mu_{0}\mathrm{H}_{\mathrm{r}}$ as a function of in-plane angle $\phi$ for $\mathrm{f}$ = 4 GHz are shown in Fig.~\ref{fig5}a. The data is fitted with the following expression, using the saturation magnetization ($\mu_{0}$$\mathrm{M}_{\mathrm{s}}$ = 1030 mT) and frequency ($\mathrm{f}$ = 4 GHz) as a fixed parameter. 
\begin{widetext}
\begin{equation}
 \mathrm{H}_{\mathrm{r}} = -\mathrm{H}_{\mathrm{k}}+\frac{3}{2}\mathrm{H}_{\mathrm{k}}\sin^2{(\phi+\Delta)}-\frac{\mathrm{M}_{\mathrm{s}}-\mathrm{H}_{\mathrm{p}}}{2}+\frac{1}{2}\left[\mathrm{H}^2_{\mathrm{k}}\sin^4{(\phi+\Delta)} +(\mathrm{M}_{\mathrm{s}}-\mathrm{H}_{\mathrm{p}})^2 +2(\mathrm{M}_{\mathrm{s}}-\mathrm{H}_{\mathrm{p}}) \mathrm{H}_{\mathrm{k}}\sin^2{(\phi+\Delta)}+\left(\frac{2\mathrm{f}}{\mu_{0}\gamma}\right)^2\right]^{\frac{1}{2}}\label{eq5}
\end{equation}
\end{widetext}
where $\mu_{0}\mathrm{H}_{\mathrm{k}}$ is the in-plane anisotropy field directed along $\theta = \phi = 0^{\circ}$ and $\mu_{0}\mathrm{H}_{\mathrm{p}}$ is the perpendicular magnetic anisotropy field lying along $\theta = 90^{\circ}$. Here $\Delta$ accounts for the offset in the magnitude of the lowest or the highest resonance field. This fitting allows us to estimate the values of  $\mu_{0}\mathrm{H}_{\mathrm{k}}$ and $\mu_{0}\mathrm{H}_{\mathrm{p}}$  anisotropies. 

In order to investigate the effects of the FM layer thickness on magnetic anisotropy, we measure the $\mu_{0}\mathrm{H}_{\mathrm{k}}$ and $\mu_{0}\mathrm{H}_{\mathrm{P}}$ as shown in Fig.~\ref{fig5}b and Fig.~\ref{fig5}c, respectively, as a function of CoFeB layer thickness for $\mathrm{f} = 4$ GHz. The error bars correspond to the maximum deviation of the anisotropy fields from the average value. It is observed that the $\mu_{0}\mathrm{H}_{\mathrm{k}}$ increases with increasing CoFeB layer thickness. On the other hand, $\mu_{0}\mathrm{H}_{\mathrm{P}}$ decreases with increasing CoFeB layer thickness. It is clear from Fig.~\ref{fig5}b and Fig.~\ref{fig5}c that the magnetic easy axis of the FM layer tilts from IP to OOP when the thickness is reduced. (For $\mathrm{f} = 6$ GHz as shown in Fig.~\ref{fig A3}b and Fig.~\ref{fig A3}c in the Appendix, we observe similar trends for $\mu_{0}\mathrm{H}_{\mathrm{k}}$ and $\mu_{0}\mathrm{H}_{\mathrm{P}}$ with varying FM layer thickness.)

A large increase in effective Gilbert damping factor is observed for the lower FM layer thickness in Pt(4 nm)/ Co$_{60}$Fe$_{20}$B$_{20}$(1.5 - 2 nm)/ Al(4 nm) stacks while a comparatively small effective Gilbert damping constant is observed for FM layers with higher thickness. Similar variation of effective Gilbert damping constant with FM layer thickness is observed for reference sample Co$_{60}$Fe$_{20}$B$_{20}$(1.5 - 2 nm)/ Al(4 nm). Experimentally observed data of effective Gilbert damping constant for both stacks are shown in the Appendix Fig.~\ref{figA2}a. Spin-pumping induced Gilbert damping contribution, $\alpha_{\mathrm{sp}} = \Delta\alpha = \alpha_{\mathrm{CoFeB/Pt}} - \alpha_{\mathrm{CoFeB}}$ (Fig.~\ref{fig 6}a) is utilised to estimate the effective spin-mixing conductance ($\mathrm{g}^{\uparrow\downarrow}_{\mathrm{eff}}$) as shown in Fig.~\ref{fig 6}b. The spin mixing conductance ($\mathrm{g}^{\uparrow\downarrow}_{\mathrm{eff}}$)determines the amount of spin current transmitted by the precessing magnetization vector of FM layer across the interface and is given by the following expression~\cite{Tser}.
\begin{eqnarray}
\mathrm{g}^{\uparrow\downarrow}_{\mathrm{eff}} =\frac{4\pi \mathrm{M}_{s}\mathrm{t}_{\mathrm{FM}}}{\mathrm{g}\mu_{B}}(\alpha_{\mathrm{CoFeB/Pt}}-\alpha_{\mathrm{CoFeB}})\label{eqn6}
\end{eqnarray}
where $\mu_{B}$ is the Bohr magneton. For CoFeB layers with lower thickness, $\alpha_{\mathrm{sp}}$ is found to be relatively high (Fig.~\ref{fig 6}a). The extracted values of($\mathrm{g}^{\uparrow\downarrow}_{\mathrm{eff}}$) decreases with the increasing thickness of the FM layer (Fig.~\ref{fig 6}b). The values of ($\mathrm{g}^{\uparrow\downarrow}_{\mathrm{eff}}$) are found to be within the range of $2.1 - 3.42 \times 10^{18}$ m$^{-2}$. These values are comparable to the reported value for Pt/ Co$_{60}$Fe$_{20}$B$_{20}$ stack~\cite{A,Ruiz}. 

The enhancement of the Gilbert damping observed in the Co$_{60}$Fe$_{20}$B$_{20}$/ Pt stacks could be interpreted in terms of the spin current injected in the Pt layer by the spin pumping mechanism at the CoFeB/Pt interface. The dissipation of spin current at the FM/NM interface by spin-flip scattering acts as an additional channel for anisotropic spin relaxation, leading to enhanced damping $\alpha_{\mathrm{sp}}$. The diffusive flow of spins in CoFeB/ Pt stacks can be described by spin current density $\mathrm{J}_{\mathrm{s}}$, evaluated using the following expression~\cite{Tser,Tser2}:
\begin{widetext}
\begin{equation}
\mathrm{J}_{\mathrm{s}} \approx \left(\frac{\mathrm{g}^{\uparrow\downarrow}_{\mathrm{eff}}\hbar}{8\pi}\right)\left(\frac{\mu_0
\mathrm{h}_{\mathrm{rf}}\gamma}{\alpha}\right)^2 \left[\frac{\mu_0 \mathrm{M}_{\mathrm{s}}\gamma+\sqrt{(\mu_0 \mathrm{M}_{\mathrm{s}}\gamma)^2+16(\pi \mathrm{f})^2}}{(\mu_0\mathrm{M}_{\mathrm{s}}\gamma)^2+16(\pi \mathrm{f})^2}\right]\left(\frac{2e}{\hbar}\right) \label{eq7}
\end{equation}
\end{widetext}
where ($\mu_{0}\mathrm{h}_{\mathrm{rf}}$) is the RF magnetic field of $1$ Oe (at 15-dBm rf power) in the strip line of the coplanar waveguide. The calculated values of $\mathrm{J}_{\mathrm{s}}$ at $\mathrm{f} = $ 4 GHz is shown in Fig.~\ref{fig 6}c. It is clearly observed that the spin-current density increases with the increasing CoFeB layer thickness. Such enhancement in spin current density provides direct evidence of the interfacial enhancement of the Gilbert damping in these CoFeB/Pt stacks.

\section{Conclusion}

In conclusion, we experimentally demonstrate the interfacial magnetic anisotropy dependent spin pumping in the CoFeB/Pt stack. The reorientation of magnetic easy axis from in-plane to out-of-plane is controlled by reducing the thickness of the CoFeB layer. The CoFeB layer with 1.5 nm thickness shows perpendicular magnetic anisotropy with negligible in plane contribution whereas CoFeB layer with 2.0 nm thickness shows significantly higher in-plane magnetic anisotropy. The anisotropy field dependent modulation of Gilbert damping factor due to spin pumping is observed. The damping modulation is co-related with the tilted magnetic easy axis of the ferromagnetic layer with reducing thickness. The change in Gilbert damping results in spin-mixing conductance ($\mathrm{g}^{\uparrow\downarrow}_{\mathrm{eff}}$) in the range of 2.1 - 3.42 $\times$10$^{18}$ m$^{-2}$. The spin-current density is also estimated and found to increase with increasing in-plane anisotropy. This is clearly related to the interfacial anisotropy dependent spin pumping and consequently spin to charge conversion. The present investigation provides important insights into the interfacial effect on spin dynamics in CoFeB/ Pt stack, crucial for spin-torque based memory applications.

\section{Acknowledgements}
MT acknowledges the MHRD, Government of India for senior Research Fellowship. SM acknowledges the Department of Science and Technology (DST), Government of India for financial support. RM would like to acknowledge the initiation grant, IIT Kanpur (IITK/PHY/2022027).

\appendix

\section{ Lorentzian fitting function for FMR derivative signal}

\begin{figure*}
\includegraphics[width=\linewidth]{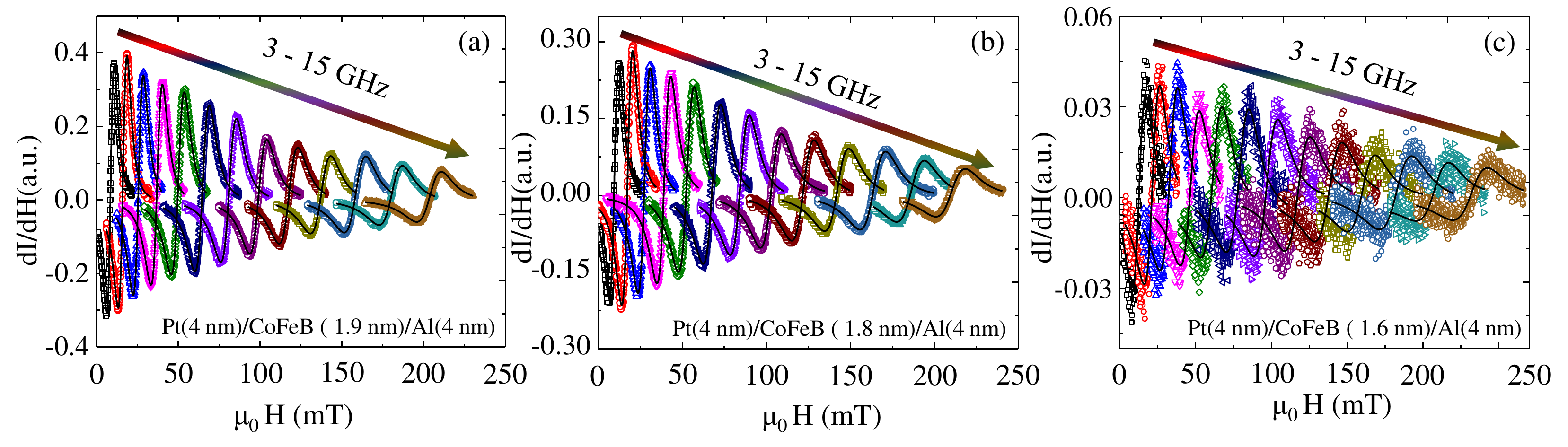}
\caption{FMR spectra recorded on a Pt(4 nm)/CoFeB($t_{\mathrm{CoFeB}}$)/Al(4 nm) stacks. Open symbols are experimental data points and black solid lines show the fitting using Eq.~A\ref{eq9}. (a) {Co$_{60}$Fe$_{20}$B$_{20}$} (1.9 nm) (b) {Co$_{60}$Fe$_{20}$B$_{20}$} (1.8 nm) (c) {Co$_{60}$Fe$_{20}$B$_{20}$} (1.6 nm)}
\label{fig A1}
\end{figure*}

Fig.~\ref{fig A1} shows the recorded FMR spectra of Pt(4 nm)/ Co$_{60}$Fe$_{20}$B$_{20}$(t$_{\mathrm{CoFeB}}$)/ Al(4 nm) stack for different CoFeB layer thickness. The FMR data is usually fitted to the standard Lorentzian fitting equation as shown in Eq.~\ref{eq1}. However, we use a modified fitting function as shown below.
\begin{widetext}
\begin{equation}
\frac{d\mathrm{I}_{\mathrm{FMR}}}{d\mathrm{H}}=4\mathrm{A}\frac{\Delta \mathrm{H}(\mathrm{H}-\mathrm{H}_{\mathrm{r}})}{(4(\mathrm{H}-\mathrm{H}_{\mathrm{r}})^2+(\Delta\nonumber \mathrm{H})^2)^2}
-\mathrm{S}\frac{(\Delta \mathrm{H})^2-4(\mathrm{H}-\mathrm{H}_{\mathrm{r}})^2}{(4(\mathrm{H}-\mathrm{H}_{\mathrm{r}})^2+(\Delta \mathrm{H})^2)^2}+ \mathrm{a}\mathrm{H}+ \mathrm{b}
\end{equation}
\end{widetext}
Here $\mathrm{a}$ and $\mathrm{b}$ are constants. The extra term $(\mathrm{a}\mathrm{H} + \mathrm{b})$ represents the small background signal which is introduced to account for any systematic increase in the diode voltage (detector) with increasing applied DC magnetic field~\cite{Celinski}.

\section{ Estimation of thickness dependent interfacial magnetic anisotropy} 

This section describes how to establish a correlation between the azimuthal angle of the magnetization vector of an in-plane magnetised film and the resonant magnetic field of FMR, which is utilised to determine the magnetic anisotropy field of perpendicular magnetic anisotropy (PMA) and in plane magnetic anisotropy (IMA). The in plane angle-dependent FMR measurement is a reliable technique for determining magnetic anisotropy in FM thin films when compared to the same calculated from magnetic hysteresis measurements~\cite{Rohit,Deka}. This is especially true when the magnetic anisotropy field is much smaller than the demagnetizing field. Moreover, to get rid of nonlinear magnetization dynamics, which changes the demagnetizing field in the sample during the FMR measurements, low input power should be utilized. This allows us to determine anisotropy fields down to a few Oersteds~\cite{Rohit,Medwal}. 
\begin{figure*}
\includegraphics[width=0.7\linewidth]{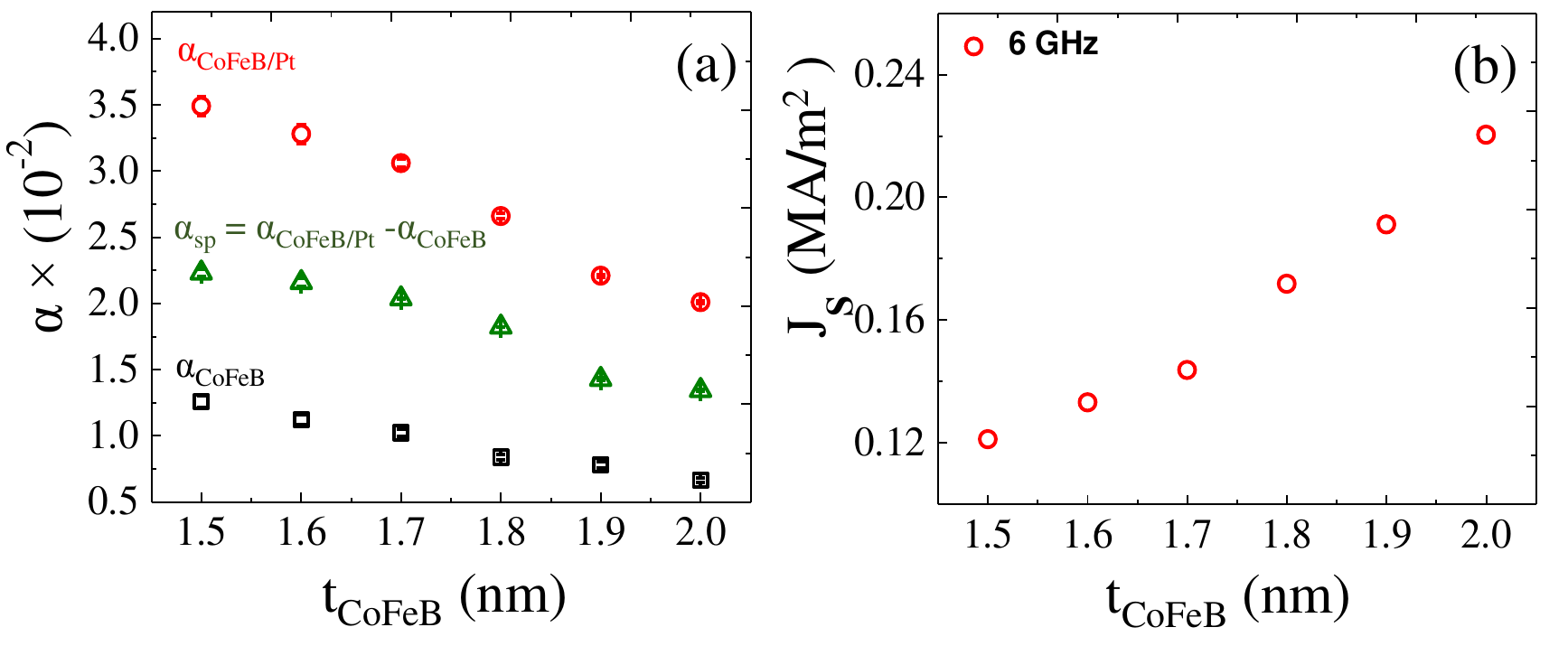}
\caption
{(a) Gilbert damping factors with respect to the thickness ($\mathrm{t}_{\mathrm{CoFeB}}$) of CoFeB layer.(b) Effective spin-current density (injected in Pt) vs $\mathrm{t}_{\mathrm{CoFeB}}$ calculated at $\mathrm{f} = 6$ GHz using Eq.~\ref{eq7}.}
\label{figA2}
\end{figure*}
\begin{figure*}
\includegraphics[width=0.8\linewidth]{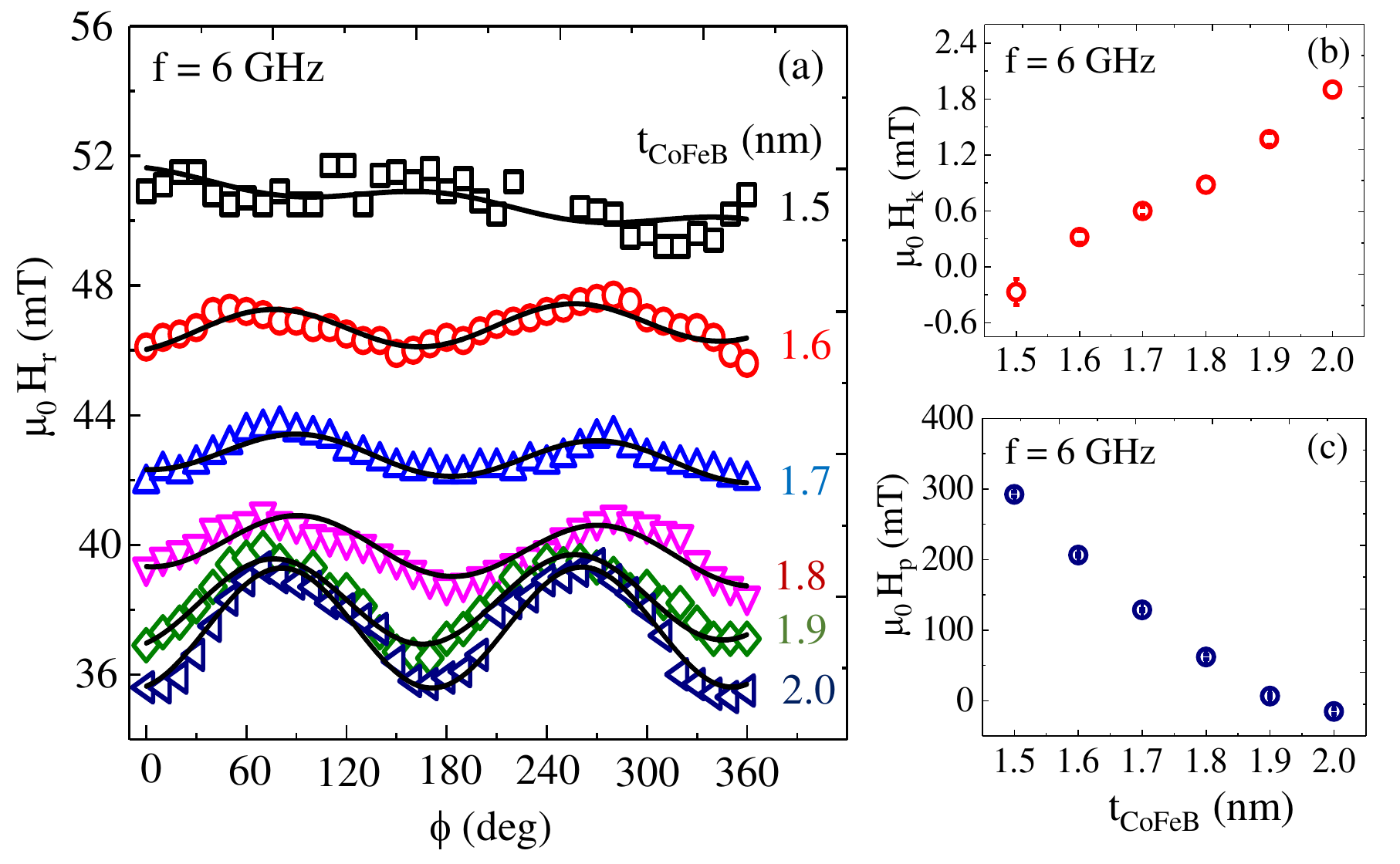}
\caption{(a) Azimuthal angle $\phi$ dependence of $\mu_{0}\mathrm{H}_{\mathrm{res}}$ for $\mathrm{f} = 6$ GHz. Sinusoidal dependence shows the presence of an in-plane anisotropy in addition to the PMA. The experimental data are fitted (solid line) with (Eq.~\ref{eq5}) to obtain $\mu_{0}\mathrm{H}_{\mathrm{k}}$ and $\mu_{0}\mathrm{H}_{\mathrm{p}}$. (b) and (c) shows the variation of in plane magnetic anisotropy $\mu_{0}\mathrm{H}_{\mathrm{k}}$ and out of plane magnetic anisotropy $\mu_{0}\mathrm{H}_{\mathrm{p}}$ vs thickness of CoFeB layer $\mathrm{t}_{\mathrm{CoFeB}}$}.
\label{fig A3}
\end{figure*}
Consider that the direction of the external magnetic field ($\mathrm{H}_{\mathrm{DC}}$) and the magnetization $\mathrm{M}$ are determined by (${\phi}$ , ${\theta}$) and (${\alpha}$, ${\beta}$), respectively, where (${\alpha}$, ${\phi}$) are the azimuthal and (${\beta}$, ${\theta}$) polar angles with respect to the film plane. When the Zeeman energy, demagnetization energy, and magnetic anisotropy energy are taken in consideration, the magnetic free energy $\mathrm{F}$ is described as follows~\cite{Gallardo}.
\begin{widetext}
\begin{equation}
 \mathrm{F} = \frac{\mu_{0}{\mathrm{M}_{\mathrm{s}}}}{2}[-2\mathrm{H}_{\mathrm{ex}}(\cos{\theta}\cos{\beta}\cos{(\alpha-\phi)+
 \sin{\theta}\sin{\beta}})+\\(\mathrm{M}_{\mathrm{s}}-\mathrm{H}_{\mathrm{p}})\sin^2{\theta}-\mathrm{H}_{\mathrm{k}}\cos^2{\theta}\cos^2{\phi}]
\end{equation}
\end{widetext}
Next we substitute $F$ in the Smit-Beljers relation~\cite{Smit} given by,
\begin{equation}
{\left(\frac{\mathrm{f}}{\gamma}\right)^2}=\frac{1}{\left(\mathrm{M}_{\mathrm{s}}\cos\theta\right)^{2}}\left[{\frac{\partial^2\mathrm{F}}{\partial\theta^2}}{\frac{\partial^2\mathrm{F}}
{\partial\phi^2}}-\left(\frac{\partial^2\mathrm{F}}{\partial\theta \partial\phi}\right)^2\right]
\label{eq9}
\end{equation}
We can obtain the free energy differentials with respect to $\theta$ and $\phi$ as follows:
\begin{widetext}
\begin{eqnarray}  
\frac{\partial^2\mathrm{F}}{\partial\mathrm{\theta}^2} & = & \mu_{0}{\mathrm{M}_{\mathrm{s}}}\left[\mathrm{H}_{\mathrm{ex}}\sin{\theta}\sin{\beta}+\mathrm{H}_{\mathrm{ex}}\cos{\beta}\cos(\alpha-\phi)\cos{\theta}-(\mathrm{M}_{\mathrm{s}}-\mathrm{H}_{\mathrm{p}})\cos{2\theta}-\mathrm{H}_{\mathrm{k}}\cos^2{\phi}\sin{2\theta}\right]\\
\frac{\partial^2\mathrm{F}}{\partial\mathrm{\phi}^2} & = & \mu_{0}\mathrm{M}_{\mathrm{s}}\left[\mathrm{H}_{\mathrm{ex}}\cos{\theta}\cos{\beta}\cos (\alpha-\phi) + \mathrm{H}_{\mathrm{k}}\cos^2{\theta}\cos{2\phi}\right]\\
\frac{\partial^2\mathrm{F}}{\partial\mathrm{\theta} \partial\mathrm{\phi}} & = & -\mu_{0}\mathrm{M}_{\mathrm{s}} \left[\mathrm{H}_{\mathrm{ex}}\sin{\theta}\cos{\beta}\sin (\alpha-\phi) + \frac{1}{2}\mathrm{H}_{\mathrm{k}}\sin{2\theta}\sin{2\phi}\right]
\end{eqnarray}
\end{widetext}
In this study, in-plane FMR measurements are carried out while $\mathrm{M}$ remains parallel to the $\mathrm{H}_{\mathrm{ex}}$ to determine the $\mathrm{H}_{\mathrm{p}}$ and $\mathrm{H}_{\mathrm{K}}$. Using $\theta = \beta = 0^{\circ}$ and $\phi = \alpha$ and substituting the resulting expression in Eq.~\ref{eq9}, we obtain the following relation:
\begin{widetext}
\begin{equation}
{\left(\frac{\mathrm{f}}{\gamma}\right)^2} = \frac{1}{\mathrm{M}_{\mathrm{s}}^{2}}\left[{\mu_{0}{\mathrm{M}_{\mathrm{s}}}}\mathrm{H}_{\mathrm{ex}}+{\mu_{0}{\mathrm{M}_{\mathrm{s}}}}\left(\mathrm{M}_{\mathrm{s}}-\mathrm{H}_{\mathrm{p}}\right)+{\mu_{0}{\mathrm{M}_{\mathrm{s}}}}\mathrm{H}_{\mathrm{k}}\cos^2{\phi}\right]
\left[{\mu_{0}{\mathrm{M}_{\mathrm{s}}}}\mathrm{H}_{\mathrm{ex}}+{\mu_{0}{\mathrm{M}_{\mathrm{s}}}}\mathrm{H}_{\mathrm{k}}\cos{2\phi}\right]
\end{equation}
\end{widetext}
Replacing $\mathrm{H}_{\mathrm{ex}}$ with the resonant magnetic field $\mathrm{H}_{\mathrm{res}}$ in the above Equation, we obtain the relation between $\mathrm{H}_{\mathrm{res}}$ and ${\phi}$ shown in Eq.~\ref{eq5}.


 For  $\mathrm{f} = 6$ GHz, we measure the $\mu_{0}\mathrm{H}_{\mathrm{k}}$ and $\mu_{0}\mathrm{H}_{\mathrm{P}}$ as a function of CoFeB layer thickness as shown in Fig.~\ref{fig A3}b and Fig.~\ref{fig A3}c, respectively. It is observed that the $\mu_{0}\mathrm{H}_{\mathrm{k}}$ increases and $\mu_{0}\mathrm{H}_{\mathrm{P}}$ decreases as the CoFeB layer thickness increases from 1.5 nm to 2.0 nm, as observed for  $\mathrm{f} = 4$ GHz (shown in  Fig.~\ref{fig5}(b,c) in the main text).

\end{document}